\documentclass[12pt]{report}

\begin{document}

\begin{center}
Abstract Submitted \\
for the APR05 Meeting of \\
The American Physical Society
\end{center}

\begin{flushright} 
Sorting Category: E10. (T)
\end{flushright}

\begin{quotation}
\begin{center}
{\bf Alternate derivation of the Ginocchio--Haxton relation $[(2j-3)/6]$}

ALBERTO ESCUDEROS and LARRY ZAMICK \\ (Rutgers University)
\end{center}

We want the number of states with total angular momentum $J=j$ for 3 identical 
particles (e.g., neutrons) in a $j$ shell. We form states $M_1>M_2>M_3$ with 
total $M=M_1+M_2+M_3$. Consider first all states with $M=j+1$. Next form states
by lowering $M_3$ by one. All such states exist because the lowest value of 
$M_3$ is $(j+1)-j-(j-1)=-j+2$. So far we have the total number of states with 
$J>j$ and $M=j$. The additional states with $M=j$ are the states with $J=j$. 
These additional states have the structure $M_1,M_2,M_2-1$ because if we try to
raise $M_3$ we get a state not allowed by the Pauli principle, namely 
$M_1,M_2,M_2$. The possible values of $M_1,M_2$ are, respectively, $j-2n$ and 
$1/2+n$, where $n=0,1,2 \cdots$. The total number of $J=j$ states is 
$N=\bar{n}+1$ (with $\bar{n}=n_{\rm max}$), while $\bar{n}$ itself is the 
number of seniority 3 states. The condition $M_1>M_2$ leads to $\bar{n} <
(2j-1)/6$ or $N<(2j+5)/6$. This is our main result. It is easy to show that 
this is the same as the Ginocchio--Haxton relation\footnote{J.N.~Ginocchio and 
W.C.~Haxton, {\it Symmetries in Science VI}, ed. by B.~Gruber and M.~Ramek, 
Plenum, New York (1993).} (see also Talmi's 1993 book) $\bar{n}=[(2j-3)/6]$, 
where $[$ $]$ means the largest integer. Since $2j$ is an odd integer, 
$(2j-1)/6$ is either $I,I-1/3$ or $I-2/3$, where $I$ is an integer. If the 
value is $I$, then $\bar{n}=[(2j-3)/6]=[I-1/3]=I-1$. It is easy to show 
agreement in the other two cases as well. The number of $J=j$ states for the 
3-particle system is equal to the number of $J=0$ states for a 4-particle 
system.
\end{quotation}

\end{document}